\documentclass[12pt,%reprint,
 amsmath,amssymb,
 aps,
]{revtex4-1}
\usepackage{color}
\usepackage[T1]{fontenc}
\usepackage{fancyhdr}
\usepackage{enumerate}
\usepackage{graphicx}% Include figure files
\usepackage{dcolumn}% Align table columns on decimal point
\usepackage{bm}% bold math

\begin{document}

\title{ Counter-term for the Palatini action of general relativity}
\author{Jingbo Wang}
\email{shuijing@mail.bnu.edu.cn}
\affiliation{College of Physics and Electronic Engineering, Xinyang Normal University, Xinyang, 464000, P. R. China}
 \date{\today}
\begin{abstract}
In this paper, we analysis the counter-term for the general relativity in the Palatini framework. The expression is valid for both the null boundary and non-null boundary. We show that final results coincide with that in Ref.\cite{pad1} which starts form the Einstein-Hilbert action.
\end{abstract}
\maketitle

\section{Introduction}
It is well known that on a spacetime with boundary, the variational principle is not well-posed for the Einstein-Hilbert action. One need to add a counter-term to the action. The usual one is the Gibbons-Hawking-York(GHY) term \cite{ghy1,ghy2} which is well-defined for space-like or time-like boundary. But this term can not be used directly to the null boundary.

In Ref.\cite{pad1} the authors tackle this problem and give a counter-term for the null boundary, and lately \cite{pad2} give a unified expression for null and non-null boundaries. The starting point is the Einstein-Hilbert action, which consider the metric as the basic variables.

There is another action--the Palatini action, which consider the connection and co-tetrad form fields as basic variables. This action is used in loop quantum gravity \cite{loop1,loop2,loop3,loop4}, which is a candidate for quantum gravity. Another advantage of this first order formula is that, when considering a black hole, the boundary degrees of freedom can be described by a SO$(1,1)$ BF theory \cite{wmz,wh1,wh2,hw4}. And this BF theory can also be applied to balck holes in Lovelock theory \cite{whl} and scalar-tensor theory \cite{wh3}. In this paper, we analysis the counter-term for the Palatini action, and find that the final results coincide with that in Ref.\cite{pad1}.

This paper is organized as follows. In section 2, the counter-term for a general boundary is obtained. In section 3, the spacelike boundary is considered and we also get the conjugate momentum. In section 4, the null boundary is considered. Our results are concluded in section 5. In appendix, we give some detail calculations. In this paper, we use the unit $c=1$, and set $\kappa=8\pi G$.

\section{The counter-term}
In this section, we deduce the counter-term for the Palatini action when the manifold $\mathcal{V}$ has a boundary $\partial\mathcal{V}$. The starting point is the Palatini action for the GR:
\begin{equation}\label{1}
    S[e,A]=-\frac{1}{2\kappa}\int_{\mathcal{V}}\Sigma_{IJ} \wedge F^{IJ},
\end{equation}
where $e^I$ are co-tetrad 1-form fields,
\begin{equation}\label{2}
    \Sigma_{IJ}=\frac{1}{2}\varepsilon_{IJKL}e^K \wedge e^L
\end{equation}
2-form fields, $A^{IJ}$ the SO$(3,1)$ connection 1-form fields, $F_{IJ}$ the curvature 2-form fields
of $A^{IJ}$, and $I,J$ indices of the Lie algebra of $\frak{so}(3,1)$.

The variation gives
\begin{equation}\label{3}\begin{split}
    \delta S[e,A]=-\frac{1}{2\kappa}\int_{\mathcal{V}} \delta \Sigma_{KL} \wedge F(A)^{KL}+d_A(\Sigma_{KL} \wedge \delta A^{KL})-d_A \Sigma_{KL}\wedge \delta A^{KL}\\=-\frac{1}{2\kappa}\int_{\mathcal{V}} \delta \Sigma_{KL} \wedge F(A)^{KL}-d_A \Sigma_{KL}\wedge \delta A^{KL}-\frac{1}{2\kappa}\int_{\partial\mathcal{V}}\Sigma_{KL} \wedge \delta A^{KL}.
\end{split}\end{equation}

In order to have a well-posed variation, one can choose the boundary condition
\begin{equation}\label{4}
 \delta A^{KL}|_{\partial\mathcal{V}}=0,
\end{equation}
or one can have another choice: since the $\Sigma_{IJ}$ fields are connected with the metric fields, we want those fields to be fixed on the boundary. To achieve this goal, a counter-term can be added to the Panatini action
\begin{equation}\label{5}
    S_{bd}=\frac{1}{2\kappa}\int_{\partial\mathcal{V}}\Sigma_{KL} \wedge A^{KL},
\end{equation}
with boundary condition
\begin{equation}\label{6}
 \delta \Sigma_{KL}\wedge A^{KL}|_{\partial\mathcal{V}}=0.
\end{equation}

Put it in another way,
\begin{equation}\label{7}
  \delta S_{\partial\mathcal{V}}=-\frac{1}{2\kappa}\int_{\partial\mathcal{V}}\delta(\Sigma_{KL} \wedge A^{KL})-\delta\Sigma_{KL} \wedge A^{KL}.
\end{equation}
\section{Spacelike boundary}
First let us consider the boundary $\partial\mathcal{V}$ to be spacelike with normal $n_c=N (dt)_c$, thus $\partial\mathcal{V}$ is the surface $t=Constant$ and $N$ is the lapse function. Choose the time gauge for the co-tetrad
\begin{equation}\label{8}
    (e^0)_a=-n_a=-N (dt)_a,
\end{equation}
then on the surface $\partial\mathcal{V}$ one have
\begin{equation}\label{9}
    e^0\triangleq 0,
\end{equation}where the symbol $\triangleq$ means the equality holds on $\partial\mathcal{V}$.

Since the connection 1-form fields are
\begin{equation}\label{10}
    A_0^{\ i}=(e_0)^c \nabla_a (e^i)_c dx^a=-(e^i)_c\nabla_a (e_0)^c dx^a,
\end{equation}
the boundary action (\ref{5}) becomes
\begin{equation}\label{11}\begin{split}
    S_{bd}=-\frac{1}{\kappa}\int_{\partial\mathcal{V}}\Sigma_{0i} \wedge A^{0i}=\frac{1}{\kappa}\int_{\partial\mathcal{V}}e^1 \wedge e^2 \wedge e^3 ( e^c_i e^a_i \nabla_a n_c )\\
    =\frac{1}{\kappa}\int_{\partial\mathcal{V}}\sqrt{h}d^3x (h^{ac} \nabla_a n_c )=\frac{1}{\kappa}\int_{\partial\mathcal{V}}\sqrt{h}d^3x(-K),
\end{split}\end{equation}
where $h_{ac}$ is the induced metric, and $K=-h^{ac} \nabla_a n_c$ is the trace of the extrinsic curvature \cite{padbook}.
This is just the usual Gibbons-Hawking-York term.
\subsection{The conjugate moment}
From the Eq.(\ref{7}) we can see that if the connection 1-form fields $A^{IJ}$ are considered to be the position variables, the $\Sigma_{IJ}$ are the conjugate momentum. But to be comparable with the results in \cite{pad1}, where the metric variables such as $h^{ab},q^{ab},l^a,k^a$ are position variables, we want to repress Eq.(\ref{6}) in terms of metric variables.

The hypersurface $\partial\mathcal{V}$ is characterized by the normal $n_a$. We assume that the variation keep the property
\begin{equation}\label{15}
    \delta n_a \propto n_a\triangleq 0.
\end{equation}
We can calculate the variation $\delta\Sigma_{KL} \wedge A^{KL}$ directly, but it is a tedious work. Instead we use the relation
\begin{equation}\label{16}
    \delta\Sigma_{KL} \wedge A^{KL}=\delta(\Sigma_{KL} \wedge A^{KL})-\Sigma_{KL} \wedge \delta A^{KL}.
\end{equation}
The first one on the right hand is known, so only the second one need to be calculated.

Since
\begin{equation}\label{17}
    \delta A_0^{\ i}=-\delta(\delta^{ij}e_j^c\nabla_a n_c dx^a),
\end{equation}
\begin{equation}\label{18}\begin{split}
    \Sigma_{KL} \wedge \delta A^{KL}=2\Sigma_{0i} \wedge \delta A^{0i}=2\Sigma_{0i} \wedge e^i e^a_j\delta(e_j^c\nabla_a n_c)\\
    =2e^1\wedge e^2 \wedge e^3 (\delta (-K)+\frac{1}{2}K_{ac}\delta h^{ac})=2\sqrt{h}d^3x (-\delta (K)+\frac{1}{2}K_{ac}\delta h^{ac}),
\end{split}\end{equation}
so the final result is
\begin{equation}\label{19}
    \delta\Sigma_{KL} \wedge A^{KL}=2d^3x(\delta(\sqrt{h}(-K))-\sqrt{h}(-\delta (K)+\frac{1}{2}K_{ac}\delta h^{ac}))\\
    =\sqrt{h}(h_{ac}K-K_{ac})\delta h^{ac}d^3x,
\end{equation}
where we use the identity
\begin{equation}\label{19a}
    \delta \sqrt{h}=-\frac{1}{2}\sqrt{h}h_{ac}\delta h^{ac}.
\end{equation}
So
\begin{equation}\label{7a}
  \delta S_{\partial\mathcal{V}}=\frac{1}{2\kappa}\int_{\partial\mathcal{V}}d^3x[\delta(2\sqrt{h}K)-\sqrt{h}(K_{ac}-h_{ac}K)\delta h^{ac}]
\end{equation}
which is the same as the Eq.(3) in Ref.\cite{pad1}.
\section{Null boundary}
Now we deal with the null case. Consider the null surface $\partial\mathcal{V}=\Delta$ with null normal $l^a$. Introduce another null vector $k^a$ with the property $l_a k^a=-1$. Assume that the null surface is given by $u=constant$, then
\begin{equation}\label{12}
    l_a=B (du)_a\triangleq 0
\end{equation} for some scalar field $B$.

The relation with the co-tetrad are $l_a=\frac{1}{\sqrt{2}}(e^0_a-e^1_a),k_a=\frac{1}{\sqrt{2}}(e^0_a+e^1_a)$. So the $\Sigma$ fields are given by
\begin{equation}\begin{split}\label{13}
    \Sigma_{01}=e^2 \wedge e^3,  \Sigma_{02}-\Sigma_{12}=-\sqrt{2}k\wedge e^3, \Sigma_{02}+\Sigma_{12}=\sqrt{2}l\wedge e^3\triangleq0,\\
    \Sigma_{23}=l \wedge k\triangleq0, \Sigma_{03}-\Sigma_{13}=\sqrt{2}k\wedge e^2, \Sigma_{03}+\Sigma_{13}=-\sqrt{2}l\wedge e^2\triangleq0.
\end{split}\end{equation}
The connection 1-form are given by
\begin{equation}\label{14}
    A_0^{\ 1}=(e_0)^c \nabla_a (e^1)_c dx^a=k_c \nabla_a l^c dx^a,\quad A_0^{\ i}+A_1^{\ i}=\sqrt{2} (e^i)_c \nabla_a l^c dx^a.
\end{equation}

Now the boundary action (\ref{5}) becomes
\begin{equation}\label{14}\begin{split}
    S_{bd}=\frac{1}{\kappa}\int_{\Delta}\Sigma_{01} \wedge A^{01}+\frac{1}{2}(\Sigma_{0i}-\Sigma_{1i})\wedge (A^{0i}-A^{1i})\\=\frac{1}{\kappa}\int_{\Delta}k \wedge e^2 \wedge e^3 (-\kappa_l- e^c_i e^a_i \nabla_a l_c )
    =\frac{1}{\kappa}\int_{\Delta}d\lambda \sqrt{q}d^2x (\kappa_l+\Theta ),
\end{split}\end{equation}
where $\kappa_l$ is the non-affinity coefficient, $\Theta$ is the expansion scalar, and we choose $k_a=-(d\lambda)_a$.
\subsection{The conjugate momentum}
Now we calculate the conjugate momentum for the null surface with the same method as in the last section. In this case, we assume
\begin{equation}\label{20}
    \delta l_a \propto l_a\triangleq 0.
\end{equation}
Then
\begin{equation}\label{21}\begin{split}
    \Sigma_{KL} \wedge \delta A^{KL}=2\Sigma_{01} \wedge \delta A^{01}+(\Sigma_{0i}-\Sigma_{1i})\wedge \delta(A^{0i}-A^{1i})\\
    =-2k \wedge e^2 \wedge e^3 (\delta \kappa_l-\delta l^a l^c \nabla_a k_c+\delta \Theta-\frac{1}{2}\Theta_{ac}\delta q^{ac}).
\end{split}\end{equation}
So
\begin{equation}\label{22}\begin{split}
    \delta\Sigma_{KL} \wedge A^{KL}=-2\delta(k\sqrt{q}d^2x (\kappa_l+\Theta ))+2k\sqrt{q}d^2x (\delta \kappa_l-\delta l^a l^c \nabla_a k_c+\delta \Theta+\frac{1}{2}\Theta_{ac}\delta q^{ac})\\
    =d\lambda \sqrt{q}d^2x ([\Theta_{ac}-q_{ac}(\kappa_l+\Theta )]\delta q^{ac}+2[k_a(\kappa_l+\Theta )+l^c \nabla_a k_c]\delta l^a)
\end{split}\end{equation}
Put it in another way,
\begin{equation}\label{22a}\begin{split}
  \delta S_{\partial\mathcal{V}}=\frac{1}{2\kappa}\int_{\partial\mathcal{V}}d\lambda d^2x[-\delta(2\sqrt{q}(\kappa_l+\Theta ))\\+\sqrt{q}[\Theta_{ac}-q_{ac}(\kappa_l+\Theta )]\delta q^{ac}+2\sqrt{q}[k_a(\kappa_l+\Theta )+l^c \nabla_a k_c]\delta l^a],
\end{split}\end{equation}
which is also the same as the Eq.(38) in Ref.\cite{pad1} besides an ignorable total derivative term.
\section{Conclusion}
We give a counter-term Eq.(\ref{5}) for the Palatini action with boundary condition Eq.(\ref{6}). Together they give a well-posed variable principle. Those expressions are same for both null boundary and non-null boundary. When we calculate those expressions for a null or non-null boundary, with some gauge condition, they give same results as those in Ref.\cite{pad1} which use the Einstein-Hilbert action.

As pointed in the introduction, the first order formulism is closely related with the loop quantum gravity, which has significate application in cosmology and black hole physics \cite{loop3}. The connection between loop quantum gravity and the "emergent gravity paradigm" \cite{egp1,egp2,egp3} is under investigation.
\appendix
\section{The detail for calculation}
In this appendix, we give the details for the calculation in section 2 and 3.
\subsection{Results for section.2}
First let's consider
\begin{equation}\label{23}
    \Sigma_{0i} \wedge \delta A^{0i}=e^2 \wedge e^3 \wedge e^1 (e^a_i\delta(e_i^c\nabla_a n_c)).
\end{equation}
Since
\begin{equation}\label{24}
 e^a_i\delta(e_i^c\nabla_a n_c)=\delta(e^a_i e_i^c\nabla_a n_c)-\delta e^a_i (e_i^c\nabla_a n_c).
\end{equation}
The first term on the right will give $h^{ac}\nabla_a n_c=-K$, and we need calculate the second term.
\begin{equation}\label{25}
 \delta e^a_i (e_i^c\nabla_a n_c)=\delta (e^a_i e_i^c)\nabla_a n_c-\delta e^c_i e_i^a\nabla_a n_c= \delta h^{ac}\nabla_a n_c-\delta e^a_i e_i^c\nabla_c n_a,
\end{equation}
so one can get
\begin{equation}\label{26}
  \delta e^a_i e_i^c(\nabla_a n_c+ \nabla_c n_a)=\delta h^{ac}\nabla_a n_c.
\end{equation}
Since we choose the time gauge condition,
\begin{equation}\label{26a}
    (e^0)_i=0\Rightarrow (e_0)^i=0,
\end{equation}and
\begin{equation}\label{27}
    dn=dN\wedge dt ,
\end{equation}so one can get
\begin{equation}\label{27a}
    \delta e^a_i e_i^c(\nabla_a n_c- \nabla_c n_a)=\delta e^a_i e_i^c(dn)_{ac}=0.
\end{equation}Thus
\begin{equation}\label{28}
  \delta e^a_i (e_i^c\nabla_a n_c)=\frac{1}{2} \delta h^{ac}\nabla_a n_c=\frac{1}{2} \delta h^{ac}(-K_{ac}-n_a a_c)=-\frac{1}{2} \delta h^{ac}K_{ac}.
\end{equation}
Combing the above results will give
\begin{equation}\label{29}\begin{split}
    \Sigma_{0i} \wedge \delta A^{0i}=e^1\wedge e^2 \wedge e^3 (\delta (-K)+\frac{1}{2}K_{ac}\delta h^{ac})=\sqrt{h}d^3x (-\delta (K)+\frac{1}{2}K_{ac}\delta h^{ac}).
\end{split}\end{equation}
\subsection{Results for section.3}
First
\begin{equation}\label{30}\begin{split}
 \Sigma_{01} \wedge \delta A^{01}=e^2 \wedge e^3 \wedge k l^a \delta(k^c \nabla_a l_c)=e^2 \wedge e^3 \wedge k [\delta(l^a k^c \nabla_a l_c)-\delta l^a (k^c \nabla_a l_c)\\=k \wedge e^2 \wedge e^3[\delta(-\kappa_l)+\delta l^a (l^c \nabla_a k_c)].
\end{split}\end{equation}
For other component,
\begin{equation}\label{31}\begin{split}
    (\Sigma_{0i}-\Sigma_{1i})\wedge \delta(A^{0i}-A^{1i})=-k \wedge e^2 \wedge e^3 e^c_i \delta(e^a_i \nabla_a l_c)=-k \wedge e^2 \wedge e^3 [\delta(e^c_i e^a_i \nabla_a l_c)-\delta e^c_i e^a_i \nabla_a l_c]\\=-k \wedge e^2 \wedge e^3 [\delta(\Theta)-\delta e^c_i e^a_i \nabla_a l_c].
\end{split}\end{equation}
With the same method as in the last subsection, one can show that
\begin{equation}\label{32}
\delta e^c_i e^a_i \nabla_a l_c=\frac{1}{2} \delta q^{ac}\nabla_a l_c= \frac{1}{2} \delta q^{ac}\Theta_{ac}.
\end{equation}
Combing all those results we can get Eq.(\ref{21}).
\acknowledgments
The author would like to thank Dr. Yu Han and Dr. Molin Liu for many help.

\bibliography{nullboundary}
\end{document}